\documentclass[%
preprint,
superscriptaddress,
amsmath,amssymb,
]{revtex4-2}
\usepackage{graphicx}
\usepackage{dcolumn}
\usepackage{bm}
\usepackage{float}
\usepackage{hyperref}
\usepackage{xcolor}
\usepackage{braket}

\begin{document}

\preprint{}

\title{Ultrastrong coupling of topologically protected edge and bulk magnetoplasmons without a dedicated cavity}

\author{F. Velli} 
\email{These two authors contributed equally to this work. \\filippo.velli@tu-dortmund.de}
\affiliation{Department of Physics, TU Dortmund University, 44227 Dortmund, Germany}
\author{P. Sai}
\email{These two authors contributed equally to this work. \\filippo.velli@tu-dortmund.de}
\affiliation{CENTERA, CEZAMAT, Warsaw University of Technology, ul. Poleczki 19, 02-822 Warsaw, Poland}
\affiliation{Institute of High Pressure Physics PAS, ul. Sokolowska 29/37, 01-142 Warsaw, Poland}
\affiliation{Faculty of Physics and CENIDE, University Duisburg-Essen, 47057 Duisburg, Germany}
\author{M. Dub}
\affiliation{CENTERA, CEZAMAT, Warsaw University of Technology, ul. Poleczki 19, 02-822 Warsaw, Poland}
\affiliation{Institute of High Pressure Physics PAS, ul. Sokolowska 29/37, 01-142 Warsaw, Poland}
\author{J. Dzian}
\affiliation{LNCMI, CNRS-UGA-UPS-INSA-EMFL, 25 rue des Martyrs, Grenoble 38000, France}
\author{F. Le Mardelé}
\affiliation{LNCMI, CNRS-UGA-UPS-INSA-EMFL, 25 rue des Martyrs, Grenoble 38000, France}
\author{M. Mittendorff}
\affiliation{Faculty of Physics and CENIDE, University Duisburg-Essen, 47057 Duisburg, Germany}
\author{M. Orlita}
\affiliation{LNCMI, CNRS-UGA-UPS-INSA-EMFL, 25 rue des Martyrs, Grenoble 38000, France}
\author{W. Knap} 
\affiliation{CENTERA, CEZAMAT, Warsaw University of Technology, ul. Poleczki 19, 02-822 Warsaw, Poland}
\author{C. Lange}
\affiliation{Department of Physics, TU Dortmund University, 44227 Dortmund, Germany}

\date \today

\begin{abstract}
In the ultrastrong light-matter coupling regime, the ability to confine the electric field to sub-wavelength scales and define an electromagnetic mode is essential. This task is typically achieved through dedicated cavities, such as THz metal resonators. Here, we instead introduce ultrastrong coupling directly between distinct magnetoplasmon excitations, leveraging solely the spatial symmetry of a confined electronic medium in the presence of a magnetic field. By patterning an \mbox{AlGaN/GaN} single quantum well hosting a two-dimensional electron gas into micrometer-scale patches, we tailor the mutual interactions between topologically protected edge and bulk magnetoplasmons, achieving coupling strengths of up to $\Omega_\mathrm{R}/\omega_\mathrm{0}$ = 0.1. The progressive breaking of rotational symmetry determines the coupling rules and leads to the emergence of energy gaps and unique chiral-mode hybridization. Our results highlight a path towards exploiting the symmetry class and topology of the magnetoplasmon wavefunction as a new parameter space for ultrastrong light-matter interaction without a dedicated cavity.
\end{abstract}

\maketitle

The role of the cavity in c-QED is to introduce a quantization of the continuum, leading to the formation of light-matter hybridized quasi-particles called cavity polaritons \cite{Ciuti2005}. In the ultrastrong coupling regime \cite{Ciuti2005,Anappara2009,Gunter2009,Scalari2012,Bayer2017,DeLiberato2017,Li2018,ParaviciniBagliani2019,FriskKockum2019,Mueller2020,Halbhuber2020,mornhinweg2021tailored,Appugliese2022,Knorr2022}, the ground state is profoundly modified by anti-resonant interactions beyond the rotating-wave approximation, leading to a finite population of virtual photons \cite{Ciuti2005,DeLiberato2017}. Their vacuum fields lead to exotic quantum effects such as the vacuum Bloch-Siegert shift \cite{Li2018}, modification of electronic transport \cite{ParaviciniBagliani2019,Appugliese2022}, control over chemical reactions \cite{Thomas2019,Dunkelberger2022}, coherent polariton-scattering \cite{Knorr2022}, and multi-wave mixing \cite{mornhinweg2021tailored}. 
Landau polaritons were the first to enter the deep-strong coupling regime~\cite{Scalari2012,Bayer2017,Mueller2020}, reaching coupling strengths of $\Omega_\mathrm{R}/\omega_\mathrm{0}=1.43$, where $\Omega_\mathrm{R}$ denotes the vacuum Rabi frequency and $\omega_0$ is the carrier frequency of light, and enabling non-adiabatic switching of the squeezed vacuum ground state~\cite{Halbhuber2020}.
Recently, the quantization of the wave vectors of light resulting from the periodic field enhancement of THz metasurfaces~\cite{Rajabali2021} has highlighted the role of magnetoplasmon excitations in Landau cavity polariton systems.
Record-strong light-matter interaction equivalent to $\Omega_\mathrm{R}/\omega_\mathrm{0}=3.19$  has been achieved by exploiting the cooperative dipole moments of multiple, highly non-resonant magnetoplasmon modes tailored by a specially designed metasurface~\cite{Mornhinweg2024b}. The role of magnetoplasmon confinement has also been underscored by controlling the spatial overlap between a structured matter component and dedicated THz metal resonators~\cite{Cortese2023, Mornhinweg2024a}. In these systems, the cavity provides an optical mode of a sufficiently high quality factor, enabling coherent, periodic energy transfer between its light field and the electronic excitation. 
However, the metal structure also imprints its near-field polarization structure onto the coupled system, masking the inherent symmetries of the polarization of the electronic excitation. Moreover, it introduces Coulomb screening, dissipation and partial modal overlap effects \cite{Khurgin2015}. 
Here, we pioneer a different approach by directly coupling distinct magnetoplasmons (MPs) through sub-wavelength confinement of the electronic medium itself, without a conventional, dedicated cavity.
\begin{figure}[b]
\includegraphics[width=0.45\textwidth]{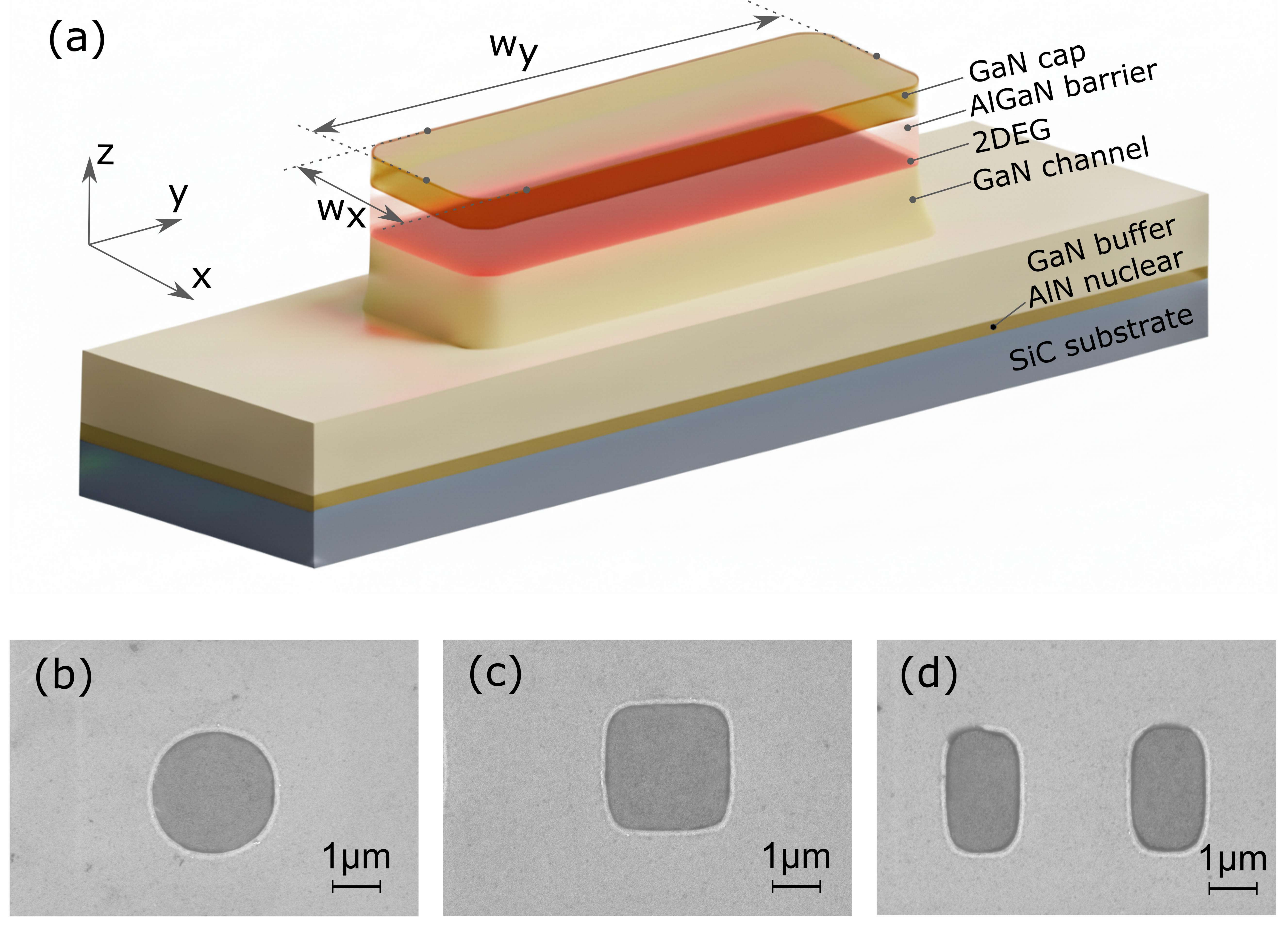}
\caption{\label{fig_1} (a) Schematic of etched AlGaN/GaN rectangular mesa with layers of the heterostructure. The GaN channel and GaN buffer are 255 nm thick, combined. (b) SEM image of etched disk structure with a diameter of $3\, \mu \mathrm{m}$. (c) Etched square patch of $3\, \mu \mathrm{m}$ side length. (d) Etched rectangular patches of $1.6\, \mu \mathrm{m}$ by $2.9\, \mu \mathrm{m}$ dimension.}
\label{fig_1}
\end{figure}
\begin{figure*}
\includegraphics[scale=1.0]{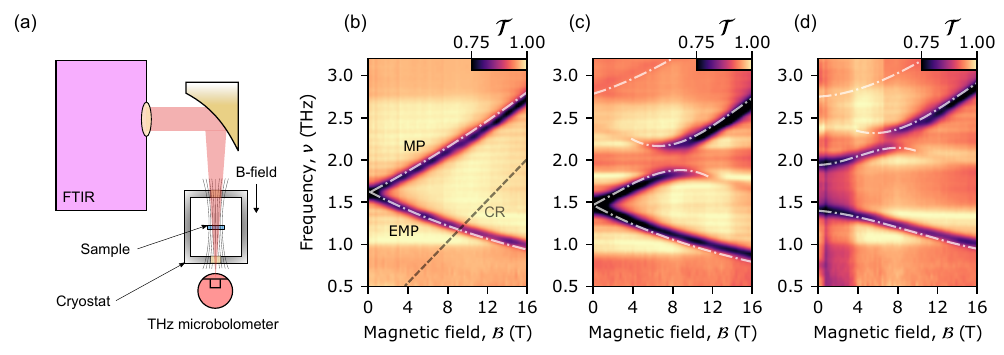}
\caption{\label{fig_2} Magnetospectroscopy of patterned GaN plasmonic structures in transmission configuration. (a) Top view of the experimental setup. The magnetic field is oriented along the beam axis and orthogonal to the plane of the 2DEG. (b)-(d) Normalized transmission spectra of disk, square, rectangular patch arrays, respectively. The overlayed white dash-dotted curves trace the transmission minima derived by our quantum model (see Fig.~\ref{fig_3}). The Rabi frequency measured in (c) is $\Omega_\mathrm{R}/\pi=0.37$ THz, at the anti-crossing point ($\omega_0/2\pi=1.9 \,\text{THz}, \, B=8.1\,$T), while in (d) it is $\Omega_\mathrm{R}/\pi=0.18$ THz at the anti-crossing point ($\omega_0/2\pi=2.2 \,\text{THz},\,B=7.5\,$T).}
\label{fig_2}
\end{figure*}
We pattern an AlGaN/GaN quantum well (QW)~\cite{Sai2023,Sai2025} hosting a high-density two-dimensional electron gas (2DEG) into micrometer-scale patches of varying shapes. Under magnetic bias, topologically distinct bulk and edge MPs~\cite{Jin2016} are formed by hybridizing plasmonic modes and cyclotron resonances (CRs) of both chiralities. For suitable symmetry conditions, they form magnetoplasmon-polaritons (MPPs) through their electrical near-field interactions and enter the ultrastrong coupling regime.
The structured 2DEG hereby takes on a dual role of defining the plasmon resonances through electronic confinement, and determining the mutual coupling of the MPs through their symmetry class. The fundamental character of the resulting MPPs can be tuned and transformed by altering the geometry of the patches: subtle changes, particularly the transition from square to rectangular geometry, open up an energy gap at zero magnetic field and modify the coupling between MPs of different chirality, in analogy with fundamental symmetry-breaking effects of quantum fields.
\\ \indent Our AlGaN/GaN heterostructure features a single quantum well hosting a 2DEG with a high carrier concentration of $n_\text{2DEG} = 9.5\times 10^{12}\,\mathrm{cm}^{-2}$ in a triangular potential, sufficient to support plasma oscillations in the THz frequency range~\cite{Dub2024}. 
The structure was grown by metal-organic vapor-phase epitaxy on a semi-insulating SiC substrate and comprises a 62-nm AlN nucleation layer, a 255-nm GaN buffer, a 20.5-nm Al$_{0.25}$Ga$_{0.75}$N barrier, and a 2.4-nm GaN cap layer.
A schematic of the layer structure is shown in Fig.~\ref{fig_1}(a). The sheet concentration of the 2DEG, which forms at the interface of the GaN and the AlGaN barrier, was first determined by capacitance-voltage measurements and subsequently verified with temperature-dependent Hall measurements. The corresponding electron Hall mobility was 12500\,cm$^2/$Vs at 10\,K, consistent with values reported for high-quality AlGaN/GaN heterostructures. Arrays of plasmonic patches with different geometries were then defined using a UV laser writer and patterned via inductively coupled plasma reactive ion etching (ICP-RIE). To electrically isolate neighboring plasmonic patches, an etch depth of 150~nm was chosen to ensure penetration of the 2DEG layer.
As a result, $3\times3$~mm$^2$-sized arrays consisting of plasmonic structures with either disks of $3\,\mu$m diameter, squares of $3\,\mu$m side length, or rectangles of $1.6\,\mu\mathrm{m}\times 2.9\,\mu$m dimension were fabricated with unit cells of $6\,\mu$m$\times6\,\mu$m, $5\,\mu$m$\times5\,\mu$m, and $4\,\mu$m$\times6\,\mu$m, respectively. These large fields strongly improve contrast in our transmission experiments while array-coupling effects remain negligible (see Supplemental Material). Representative scanning electron microscopy (SEM) images are presented in Fig.~\ref{fig_1}(b)-(d).
\\ \indent Magnetospectroscopy in transmission configuration was performed using a vacuum Fourier-transform infrared (FTIR) spectrometer equipped with a cryogenic bolometer detector positioned beneath the sample chamber [Fig.~\ref{fig_2}(a)]. The Hg-arc lamp light source produces incoherent, unpolarized radiation, and correspondingly, the transmission spectra are a superposition of a complete basis set of orthogonal polarizations. The beam measured $\lesssim 3$\,mm in diameter, and to ensure that light impinged only on the patterned area, the sample was mounted on a metallic aperture. Measurements were conducted at $4.2\,$K in magnetic fields of up to $16\,$T with a step size of $0.25\,$T and a spectral resolution of $2.0$ cm$^{-1}\,\hat{=}\,0.06$~THz. An effective electron mass of $m^* = 0.24\, m_e$ was determined from the magnetic-field dependence $\omega_\mathrm{c}(B)=eB/m^*$ measured on an unpatterned AlGaN/GaN reference sample, where $e$ and $m_e$ are the elementary charge and free electron mass, respectively.
\\ \indent The transmission spectra for disk, square, and rectangular geometries are summarized in Fig.~\ref{fig_2}. To isolate the resonant features, each spectrum is normalized to the mean transmission averaged over the full magnetic field range.
For the disk geometry [Fig.~\ref{fig_2}(b)], two modes corresponding to the first-order bulk (MP) and edge magnetoplasmon excitations (EMP) \cite{Fetter1986} are observed. They are degenerate in frequency at $B = 0$\,T and $\omega_\mathrm{p}/2\pi = 1.6$\,THz, the fundamental plasmon frequency. Since the disk belongs to the $U(1)/SO(2)$ rotational symmetry group, angular momentum is conserved. As a consequence, the magnetoplasmons are characterized by a well-defined azimuthal quantum number $l$ and are strictly orthogonal \cite{Shikin1991}.
With increasing magnetic field, the frequency of the bulk mode increases and asymptotically approaches the bare cyclotron resonance, while the EMP mode tends toward zero~\cite{Allen1983}:
\begin{align}\label{eq1}
\omega_{\text{MP}_{\pm}}=\pm\frac{\omega_\mathrm{c}}{2}+\sqrt{\omega_\mathrm{p}^2+\left(\frac{\omega_\mathrm{c}^2}{2}\right)}.
\end{align}
No higher-order modes are evident within the observed spectral range, indicating that the oscillator strength is concentrated in the fundamental mode $l=1$ \cite{Fetter1986}. 
Upon reducing the symmetry to the discrete rotation group $Z_4$ of a square [Fig.~\ref{fig_2}(c)], the rotational invariance is broken, permitting coupling between bulk and edge magnetoplasmons of different order (e.g., $\Delta l = 1$). The fundamental EMP mode exhibits comparable behavior to the disk. The frequency of the fundamental bulk mode rises with increasing $B$ field, whereas that of the second-order EMP mode of opposite chirality, starting near $2.4\,\mathrm{THz}$ for $B = 0\,\mathrm{T}$, decreases. As these frequencies converge, the modes undergo an anti-crossing to form a MPP doublet. The two MPP branches have a minimum separation of $\Omega_\mathrm{R}/\pi=0.37\,\mathrm{THz}$, corresponding to a normalized coupling strength of $\Omega_\mathrm{R}/\omega_\mathrm{0} \approx 0.1$. 
An additional mode emerging near 2.8\,THz at $B=0$\,T is attributed to the axisymmetric MP with $l=0$~\cite{Zagorodnev2023}.
\\ \indent While the hybridization of higher-order magnetoplasmon modes has been observed in microwave spectra of square patches \cite{Zarezin2023}, here we observe an exceptionally high vacuum Rabi splitting $2\Omega_\mathrm{R}$ attributable to the very large carrier concentration of our GaN QW and the excellent modal overlap at THz frequencies caused by our significantly smaller, micrometer-sized structures. Furthermore, this ultrastrong coupling is directly tied to the symmetries of the excitations of the structured matter component, which are otherwise masked by projection onto the modes of dedicated metal cavities.
The topological protection of the EMP modes \cite{Jin2016} plays a key role in their robustness to edge imperfections of the samples, which manifests by comparably low linewidths of the MPP's. The edge magnetoplasmon indeed has unbreakable particle-hole conjugation symmetry and broken time-reversal symmetry, and therefore belongs to the class-D topological phase like a Majorana wire, i.e., the current flows unidirectionally around the edges depending on the sign of the magnetic field \cite{Chiu2016}.
\begin{figure*}
\includegraphics[scale=1.0]{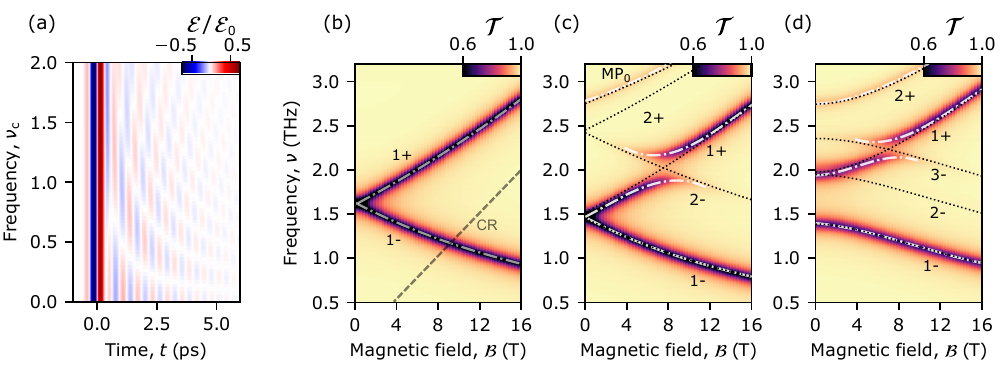}
\caption{\label{fig_3}Results of the quantum model. (a) Transmitted electric field in the case of disk magnetoplasmons. The Gaussian pulse is visible near $t=0$. (b)-(d) Fourier-transform of time-domain data for disk, square, and rectangular geometries, respectively, with the cyclotron resonance shown in panel (b) as a dashed line, for reference. Both linear polarizations are calculated individually and then superimposed. The black dotted lines represent the bare MP resonances with the notation from Eq.\,\ref{eq2}, while the white, dashed-dotted curves trace the coupled modes.}
\end{figure*}
\\ \indent Further reducing the symmetry to the $Z_2$ invariance of the rectangular patches [Fig.~\ref{fig_2}(d)], we observe the opening of a gap of $\approx 0.6$\,THz between the two lowest MPP branches, at $B = 0\,\mathrm{T}$. This zero-field gap corresponds to the intrinsic energy difference between the $x$ and $y-$polarized plasmons of the rectangular structure. When the magnetic field is applied, the lowest-energy plasmon excitation (along the longer side) hybridizes with the cyclotron resonance of opposite handedness, forming a unique edge-state MPP:
in a notable departure from the geometries discussed above, the slope $\frac{\partial}{\partial B} \omega(B)$ of their magnetic-field dependence vanishes for $B\rightarrow0$.
At $B=7.5\,\mathrm{T}$, we observe MPPs from hybridization of the lower- and higher-order MP modes as before, with a Rabi frequency of $\Omega_\mathrm{R}/\pi=0.18$\,THz.
\\ \indent This characteristic of the Landau-quantized 2DEG provides a new framework for manipulating light-matter hybridized modes. Recent studies in the field have focused on systems where the magnetoplasmons are fully screened by an electrically biased metal gate which allows to tune the electron concentration for transport experiments~\cite{Rodionov2023,Muravev2020,Zagorodnev2023}. The gate however alters the response of the 2DEG by screening the electric near-field and enforcing a linear plasmon dispersion in the long-wavelength limit. Removing the metal cavity exposes the topological properties of the electronic excitation, and the geometry via the symmetry class controls the energy gap and mode curvature in dependence on the magnetic bias.
\\ \indent To elucidate the fundamental physics governing the formation of  MPPs, we employ a quantum model of the MP interactions derived from our previous work~\cite{Mornhinweg2024b}. 
The Hamiltonian reads:
\begin{align}
\label{eq2}
\hat{H}=&\sum_{n, \sigma}\hbar\omega_{n, \sigma}^{\text{MP}}\left(B\right)\hat{b}_{n, \sigma}^\dagger\hat{b}_{n, \sigma}\,+\nonumber\\&
\sum_{n, \sigma; n', \sigma'}\hbar\Omega_\mathrm{R}^{n, \sigma; n', \sigma'}\left(\hat{b}_{n, \sigma}^\dagger+\hat{b}_{n, \sigma}\right)\left(\hat{b}_{n', \sigma'}^\dagger+\hat{b}_{n', \sigma'}\right)+\nonumber\\&\sum_{n,\sigma} \mathcal{E}_\mathrm{in}(t)\mu_{n, \sigma}\left(\hat{b}_{n, \sigma} + \hat{b}^\dagger_{n, \sigma}\right).
\end{align}
The first sum describes the bare MP resonances, where $\hat{b}_{n, \sigma}$ is the bosonic annihilation operator for the $n$-th MP mode with a frequency of $\omega_n$ and a chirality $\sigma\in\{-,+\}$. The second sum describes the coupling quantified by a matrix of vacuum Rabi frequencies $\Omega_\mathrm{R}^{n, \sigma; n', \sigma'}$, and the last sum implements the interaction with the classical coherent far field $\mathcal{E}_\mathrm{in}(t)$ coupled by corresponding dipole moments $\mu_{n, \sigma}$. We neglect the diamagnetic blue shift.
\\ \indent For disks and square patches, the dependence of the MP frequencies $\omega_{n, \sigma}^\mathrm{MP}$ on the cyclotron frequency is well described by Eq.~\ref{eq1}, which is in first approximation valid for each plasmon resonance of order $n$, and links to the denominations of bulk (1,$+$) and edge (1,$-$) MPs~\cite{Fetter1986}. These solutions are, however, geometry-specific and not generally valid. To obtain an equivalent relationship for the rectangular geometry, we apply a dynamic Drude-Lorentz model with different resonance frequencies $\omega_{n,x}$ and $\omega_{n,y}$ for the quantized $x$ and $y$ plasmons of order $n$. 
The polarizability tensor $\alpha_{ij}$ links the polarization $\mathbf{p}(\omega)=\alpha_{ij}\mathbf{E}(\omega)$ of a single oscillator to the electric field $\mathbf{E}(\omega)$ and defines the complex dielectric permittivity tensor $\varepsilon_{ij}(\omega)=1+n_\text{2DEG}\alpha_{ij}(\omega)$, where $i,j \in \{x, y\}$ (see Supplemental Material for a detailed derivation):
\begin{align}\label{eq3}
&\varepsilon_{ij}\left(\omega\right)-1=\nonumber\\&\frac{e^2n_{\text{2DEG}}}{m^\ast\varepsilon_0}\frac{-1}{\left(\omega^2-\omega_{n,x}^2+i\omega\gamma\right)\left(\omega^2-\omega_{n,y}^2+i\omega\gamma\right)-\omega_\mathrm{c}^2\omega^2} \nonumber\\
&\times\left(\begin{matrix}\omega_{n,y}^2-\omega^2+i\omega\gamma&-i\omega_\mathrm{c}\omega\\+i\omega_\mathrm{c}\omega&\omega_{n,x}^2-\omega^2+i\omega\gamma\\\end{matrix}\right).
\end{align}
Taking the high-frequency limit where damping is small ($\gamma\ll\omega$) and solving for the poles in Eq.~\ref{eq3}, we obtain the MP resonance frequencies for a given order $n$ as a function of the magnetic field:
\begin{align}
\label{eq4}
\omega_{\text{MP},n{\pm}}^2=&\frac{1}{2}\left(\omega_{n,x}^2+\omega_{n,y}^2+\omega_\mathrm{c}^2\right)\nonumber\\&\pm\left[\frac{1}{4}\left(\omega_{n,x}^2+\omega_{n,y}^2+\omega_\mathrm{c}^2\right)^2-\omega_{n,x}^2\omega_{n,y}^2\right]^{\frac{1}{2}}.
\end{align}
For $\omega_\mathrm{c}\rightarrow0$, the solutions converge to the bare plasmon frequencies $\omega_{n,x}$ and $\omega_{n,y}$. In addition, for degeneracy ($\omega_{n,x}=\omega_{n,y}$), Eq.~\ref{eq4} reduces to the square patch case (Eq. ~\ref{eq1}).
We implement this frequency dependence in our Hamiltonian and use vacuum Rabi frequencies $\Omega_\mathrm{R}^{n, \sigma;n',\sigma'}$ extracted from the experimental data at the anticrossing point. 
Including all constituent modes up to $n=3$, a total of six non-zero coupling terms populate the Rabi matrix.
The operator dynamics are derived from the Heisenberg equations of motion, and incoherent losses are implemented by damping rates $\gamma_{n,\sigma}$, such that $\frac{d}{dt} \langle \hat{b}_{n,\sigma} \rangle = -i \langle [\hat{b}_{n,\sigma}, \hat{{H}}] \rangle - \gamma_{n,\sigma} \langle \hat{b}_{n,\sigma} \rangle$.
The $\gamma_{n,\sigma}:=\gamma$ are chosen equal for all MPs and set to match the experimentally observed linewidth of FWHM $\approx 0.13$\,THz.
The incident field is represented by a single-cycle Gaussian waveform with a spectrum covering 0.1\,THz to 5~THz.
We compute the time evolution of the transmitted field by summing up all MP polarization terms and including the incident waveform, obtaining
\begin{align}\label{eq5}
\mathcal{E}_{\text{out}}(t) = \mathcal{E}_{\text{in}}(t) - \text{Re} \left[ \frac{d}{dt} \sum_{{n,\sigma}} \mu_{n,\sigma} \langle \hat{b}_{n,\sigma}(t) \rangle \right].
\end{align} 
The calculations are performed for each $B$ field and for both input polarizations. 
The transmission spectra are obtained via Fourier transformation, normalized to the spectrum of the incident pulse, and subsequently combined to account for the unpolarized experimental source. 
The results [Fig.~\ref{fig_3}(b)-(d)] accurately reproduce the experimental data, including the position of the MPP anticrossings [Fig.~\ref{fig_3}(c),(d)] and the gap for the rectangular structure [Fig.~\ref{fig_3}(d)].
For ease of comparison, dashed transparent lines trace the calculated MPP modes. The same lines are superimposed on the experimental data in Fig.~\ref{fig_2}(b)-(d), for comparison.
\begin{figure}[b]
\includegraphics{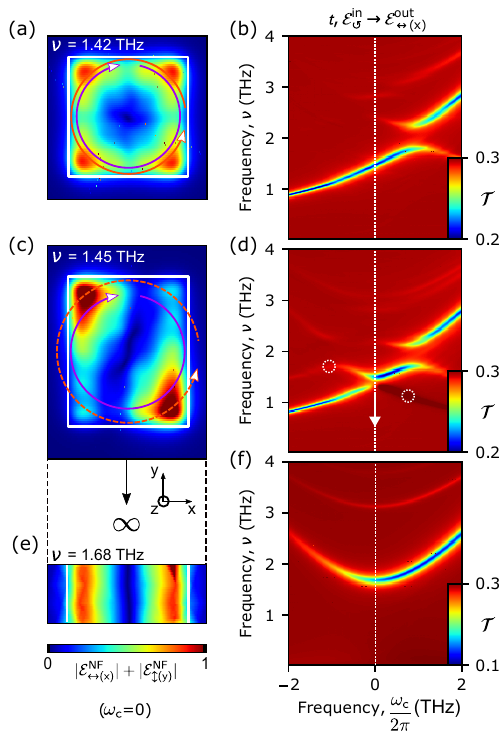}
\caption{\label{fig_4} FEM simulations for right-circularly polarized incident light. (a),(c),(e) Normalized envelope of the electric near field for the square, rectangle and stripe measuring $2.8\mu\mathrm{m}\times2.8\,\mu\mathrm{m}$, $2.8\,\mu\mathrm{m}\times3.4\,\mu\mathrm{m}$, and $2.8\,\mu\mathrm{m}$ width, respectively. The shapes are outlined in white. Collective cyclotron motion of opposite handedness is depicted by the purple and orange circular arrow shapes. (b),(d),(f) Corresponding far-field transmission spectra of the component $\mathcal{E}_x$. The white dotted circles mark MPP branches with clockwise CR hybridization.}
\end{figure}
\\ \indent To corroborate our results, we perform finite-element-method (FEM) simulations of our structures with the QW implemented as a gyrotropic medium~\cite{Bayer2017} (see Supplemental Material). 
For zero magnetic bias and linearly polarized light, the transmission of the rectangular patch reveals a fan of plasmon resonances with $\omega_{1,x}/2\pi=1.95 $\,THz and $\omega_{1,y}/2\pi=1.42$\,THz as the two fundamental dipole modes for the different confinement lengths. The subsequent plasmon resonances $(\omega_{2,x},\omega_{3,x})$ and $(\omega_{2,y},\omega_{3,y})$ are compatible with the scaling law $\omega_{n,(x,y)}\propto\sqrt{q_{n,(x,y)}}=\sqrt{\frac{n_{x,y}\pi}{L_{x,y}}}$ \cite{ando1982}. 
In addition, we open up a more intuitive understanding of the symmetry-breaking that governs the magnetic field dependence of the MPPs. 
To this end, we perform FEM simulations of the spatial near-field distribution as well as the corresponding transmission spectra, for right-circularly polarized incident light. 
In an extended system, this component couples only to the CR of corresponding chirality \cite{Fetter1986}. 
The electric near-field envelope ($\left|\mathcal{E}_x\right| + \left|\mathcal{E}_y\right|$) of the square is four-fold symmetric, independent of the handedness of the incident field [Fig.~\ref{fig_4}(a)]. 
For negative values of the cyclotron frequency, $\omega_\mathrm{c} < 0$, which correspond to a flipped magnetic bias field $B$ with respect to the cases discussed so far, only the EMP mode contributes to the transmission spectra [Fig.~\ref{fig_4}(b)], in analogy to the case of a disk-shaped structure. 
When $\omega_\mathrm{c} > 0$ instead, the hybridization of the first-order bulk and second-order EMP mode from Fig.~\ref{fig_2}(c) is recovered. 
Transitioning to the rectangular geometry [Fig.~\ref{fig_4}(c),(d)] introduces a lower-energy plasmon mode along the increased confinement length in $y$ direction, resulting in a two-fold symmetric pattern of the near field, at $B=0$. 
For a non-zero magnetic bias, this $Z_2$ symmetry lifts the degeneracy of the MPs and consequently, MPP branches associated with the opposite handedness of the CR emerge in the spectrum (white dotted circles in Fig.~\ref{fig_4}(d)). 
This crossover through $B = 0$ with a vanishing slope reflects a definitive flip in the chirality of the underlying edge modes. In the limit of vanishing magnetic field, the classical cyclotron radius tends to infinity; consequently, the collective charge motion of a single chiral edge current cannot induce plasma oscillations purely along the long side of the rectangle. Thus, both edge MPs contribute to the singular point at $B=0$, hybridizing into a unique MPP. Above the energy gap, the hybridization of the two lowest-order bulk MPs forms the characteristic S-shaped mode. 
By further extending the patch in the $y$ direction, the lowest-order MPP mode softens, eventually vanishing in the limit of an infinite stripe [Fig.~\ref{fig_4}(e),(f)]. 
Concurrently, the coupling strength and anticrossing signatures between higher-order modes diminish until the relationship $\omega_{\text{MP}}^2=\omega_p^2+\omega_\mathrm{c}^2$ is recovered and the resonances merge into an effectively one-dimensional plasmon hybridizing with the CR \cite{Kushwaha2001}.
\\ \indent In conclusion, we have observed ultrastrongly coupled magnetoplasmon-polaritons in nanostructured patches of AlGaN/GaN single quantum wells. By progressively breaking the rotational symmetry—transitioning from circular to square and ultimately to rectangular geometries—we have demonstrated how the symmetry and topological class dictates the mutual coupling between bulk and edge magnetoplasmons. Our tailored quantum model reproduces these findings, highlighting the emergence of energy gaps at zero magnetic field and chiral mode hybridizations. This establishes the geometric structuring of the electronic medium as a fundamental tool for engineering ultrastrong light-matter interactions without a dedicated cavity \cite{Canales2021}.
Future structures with engineered mode chirality are particularly interesting for excitation with circularly polarized light.
These investigations are of significant relevance for fundamental studies of cavity QED, where symmetry and topological protection are assuming an ever more prominent role, as in polaritonic systems with Dirac dispersions or plasmons in materials with inherently non-trivial topology such as topological insulators. Moreover, by transfer to a variety of promising material platforms, the realization of novel light-emitting or light-detection devices for circular polarization states is tangible.

\begin{acknowledgments}
We thank Joshua Mornhinweg for the helpful discussions regarding the quantum model, Achille Mauri for the insights on the symmetry classes, and Vincent Latko and Michael Solbach for technical assistance. The work was supported by the German Research Foundation through project B08 of TRR 142, B09 of SFB 1242, and by the European Union through the ERC Advanced Grant TERAPLASM (No.~101053716). Views and opinions expressed are, however, those of the authors only and do not necessarily reflect those of the European Union or the European Research Council Executive Agency. Neither the European Union nor the granting authority can be held responsible for them. The work was also supported by the Center for Terahertz Research and Applications (CENTERA2) project (FENG.02.01-IP.05-T004/23) carried out within the International Research Agendas program of the Foundation for Polish Science, co-financed by the European Union under European Funds for a Smart Economy Programme. The authors moreover acknowledge the support of the LNCMI-CNRS in Grenoble, a member of the European Magnetic Field Laboratory (EMFL).
\end{acknowledgments}

\bibliography{References}

\clearpage
\begin{center}
\textbf{\large Supplemental Material}
\end{center}
\vspace{2mm}

\section{Calculation of magnetic field dependence}
The relation between the resonance frequency of confined magnetoplasmons (MP) and the static magnetic bias field depends on the geometry of the patch. For the square and disk gemoetries, corresponding formulae have been derived~\cite{Fetter1986}. In the following, we employ a dynamic Drude-Lorentz formalism to derive the relationship for the rectangular patch  discussed in the main manuscript. 
We consider a 2D electron gas in the $(x,y)$ plane with a constant magnetic field $B$ perpendicular to the plane. Different MP resonance frequencies are assumed for the $x-$ and $y$-polarized plasmon resonances. 
Assuming a scalar effective mass $m^*$, the coupled equations of motion for a single oscillator read:
\begin{align*}
    \ddot{x} &= -\gamma \dot{x} - \omega_{n,x}^2 x - \frac{eB}{m^*} \dot{y} - \frac{e}{m^*} E_x \\
    \ddot{y} &= -\gamma \dot{y} - \omega_{n,y}^2 y + \frac{eB}{m^*} \dot{x} - \frac{e}{m^*} E_y
\end{align*}
Here, $\gamma$ is the damping rate and $E_x, E_y$ are the in-plane components of the incident light field. 
For a stationary solution, a temporal dependence of $E_{x,y}\propto e^{-i\omega t}$ is employed.
For the oscillator, the ansatz
\begin{align*}
x&=x_0 e^{-i\omega t}
\end{align*}
leads to
\begin{align*}
    x(\omega_{n,x}^2 - \omega^2 + i\omega\gamma) - i\omega_c \omega y &= -\frac{e}{m^*} E_x \\
    y(\omega_{n,y}^2 - \omega^2 + i\omega\gamma) + i\omega_c \omega x &= -\frac{e}{m^*} E_y
\end{align*}
Here, $\omega_c = eB/m^*$ is the cyclotron frequency. The system can be written in a more compact form as:
\begin{equation*}
    \alpha_{ij}^{-1} \begin{pmatrix} x \\ y \end{pmatrix} = \frac{e}{m^*} \begin{pmatrix} E_x \\ E_y \end{pmatrix} 
\end{equation*}
where $\alpha_{ij}^{-1}$ is the inverse of the polarizability tensor for a single dipole $p(\omega) = \alpha_{ij} E(\omega)$:
\begin{equation*}
    \alpha_{ij}^{-1} = \begin{pmatrix} 
    \omega_{n,x}^2 - \omega^2 + i\omega\gamma & -i\omega_c \omega \\ 
    +i\omega_c \omega & \omega_{n,y}^2 - \omega^2 + i\omega\gamma 
    \end{pmatrix}
\end{equation*}

The macroscopic polarization can be written by the 2DEG density $n_\text{2DEG}$ and using the relations for the macroscopic displacement field and the conductivity,
\begin{equation*}
    D = \epsilon(\omega)E = \epsilon_0 E + P = (1 + n_{\text{2DEG}} \alpha_{ij}(\omega))E, \quad \sigma(\omega) = -i\omega(\epsilon(\omega) - 1)\mbox{,}
\end{equation*}
we obtain for the complex magnetoconductivity after matrix inversion:
\begin{align*}
    \sigma_{ij}(\omega) &= \frac{n_s e^2}{m^* \epsilon_0} \frac{i\omega}{(\omega^2 - \omega_{n,x}^2 + i\omega\gamma)(\omega^2 - \omega_{n,y}^2 + i\omega\gamma) - \omega_\mathrm{c}^2 \omega^2} \\
    &\times \begin{pmatrix} 
    \omega_{n,y}^2 - \omega^2 + i\omega\gamma & -i\omega_\mathrm{c} \omega \\ 
    +i\omega_\mathrm{c} \omega & \omega_{n,x}^2 - \omega^2 + i\omega\gamma 
    \end{pmatrix}\mbox{.}
\end{align*}

In the high-frequency limit ($\omega/\gamma \gg 1$), the plasmon frequencies for each order $n$ have the form $\omega_{n,x;n,y} \propto \sqrt{q_{n,x,n,y}} = \sqrt{\frac{n_{x,y}\pi}{L_{x,y}}}$. It is important to highlight the dependence on the square root of the plasmon order $n$ and the inverse square root of the quantization dimensions $L_{x,y}$.

Solving for the poles in the conductivity expression, we can obtain the magnetoplasmon resonance frequencies of order $n$ as a function of the magnetic field:
\begin{equation*}
    \omega_{n\pm}^2 = \frac{1}{2} \left( \omega_{n,x}^2 + \omega_{n,y}^2 + \omega_\mathrm{c}^2 \pm \sqrt{(\omega_{n,x}^2 + \omega_{n,y}^2 + \omega_\mathrm{c}^2)^2 - 4\omega_{n,x}^2 \omega_{n,y}^2} \right)\mbox{.}
\end{equation*}

For $\omega_\mathrm{c} \to 0$, the solutions converge to the bare plasmon frequencies. These analytical solutions for the bare magnetoplasmon modes in rectangular geometry differ substantially from the well-known solutions for circular and square patches.
In particular, the derivative $d\omega/d\omega_\mathrm{c} \to 0$ at zero magnetic field, and for $\omega_{n,x}^2 = \omega_{n,y}^2$, the solutions match the ones of the square patches.

\vspace{1mm}
\normalsize%
\subsection{Quantum Model}%
The Hamiltonian from the main manuscript reads:
\begin{align}\label{eq2}
\hat{H}=&\sum_{n, \sigma}\hbar\omega_{n, \sigma}^{\text{MP}}\left(B\right)\hat{b}_{n, \sigma}^\dagger\hat{b}_{n, \sigma}\,+\nonumber
\sum_{n, \sigma; n', \sigma'}\hbar\Omega_\mathrm{R}^{n, \sigma; n', \sigma'}\left(\hat{b}_{n, \sigma}^\dagger+\hat{b}_{n, \sigma}\right)\left(\hat{b}_{n', \sigma'}^\dagger+\hat{b}_{n', \sigma'}\right)+\nonumber\sum_{n,\sigma} \mathcal{E}(t)\mu_{n, \sigma}\left(\hat{b}_{n, \sigma} + \hat{b}^\dagger_{n, \sigma}\right).
\end{align}
Here we briefly discuss the notation and non-zero terms for the Rabi coupling matrix. The index pairs $(n,\sigma)$, with $n\in(1,s) $, where $s$ is the highest-order plasmon resonance included for each specific geometry, and $\sigma\in\{-,+\}$, yield matrix elements $\braket{n',\sigma'|n,\sigma}\neq0$ for $\sigma'\neq\sigma$, $n'> n$. For the disk geometry the matrix is trivially zero since only two modes $\ket{1,-},\ket{1,+}$ are included. For the square case, given $s=2$ the matrix has two non-zero terms and is 
\begin{equation}
\Omega_{R}^{n,\sigma;n',\sigma'} = 2\pi \times 10^{12} 
\begin{array}{cccc} 
\scriptstyle{1- \: 1+ \  2- \ \ 2+ } \\ \hline 
\left( \begin{matrix} 
0 & 0 & 0 & 0 \\ 
0 & 0 & 0.2 & 0 \\ 
0 & 0.2 & 0 & 0 \\ 
0 & 0 & 0 & 0 

\end{matrix} \right) 
\end{array} \quad \text{s}^{-1}\mbox{.}
\end{equation}
The matrix for the rectangular case with modes up to $s=3$ is
\vspace{2mm}
\begin{equation}
\Omega_{R}^{n,\sigma;n',\sigma'} = 2\pi \times 10^{12} 
\begin{array}{cccccc} 
\scriptstyle{1- \ \ 1+ \quad \ 2- \quad \ 2+ \quad \ 3- \quad 3+} \\ \hline 
\left( \begin{matrix} 
0 & 0 & 0 & 0 & 0 & 0 \\ 
0 & 0 & 0.03 & 0 & 0.1 & 0 \\ 
0 & 0.03 & 0 & 0 & 0 & 0 \\ 
0 & 0 & 0 & 0 & 0.03 & 0 \\ 
0 & 0.1 & 0 & 0.03 & 0 & 0 \\ 
0 & 0 & 0 & 0 & 0 & 0 
\end{matrix} \right) 
\end{array} \quad \text{s}^{-1}\mbox{.}
\end{equation}

\vspace{2mm}
The coupling to the external field is controlled by the parameters $\mu_{n,\sigma}$ describing the dipole coupling. 
They are tuned to match the normalized transmission from the experimental data. For the fundamental modes $\mu_{1,\mp}=0.8$, while for the higher modes $\mu_{2,\mp}=0.1$, and $\mu_{3,\mp}=0.02$. It is important to note here that there is a coincidental degeneracy of the modes $\ket{1,+},\ket{2,-}$ at $\omega_\mathrm{c}=0$, since the bare plasmon energies $\omega_{1,x}/2\pi=1.95$\,THz and $\omega_{2,y}/2\pi=2.02$\,THz are within the experimental linewidth.

\section{FEM simulations}

\begin{figure}[h!]
    \centering
    \includegraphics[width=0.33\textwidth]{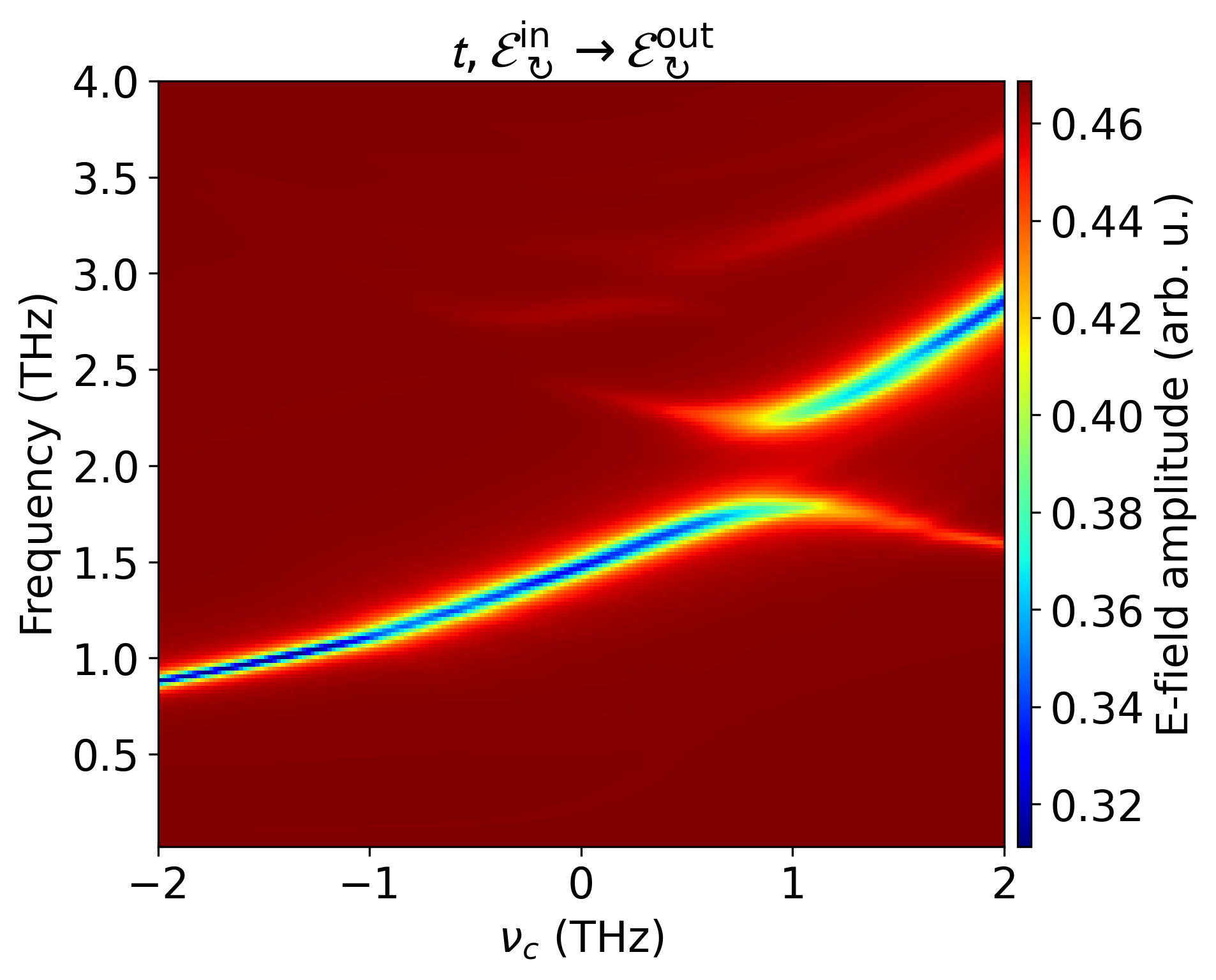}%
    \includegraphics[width=0.33\textwidth]{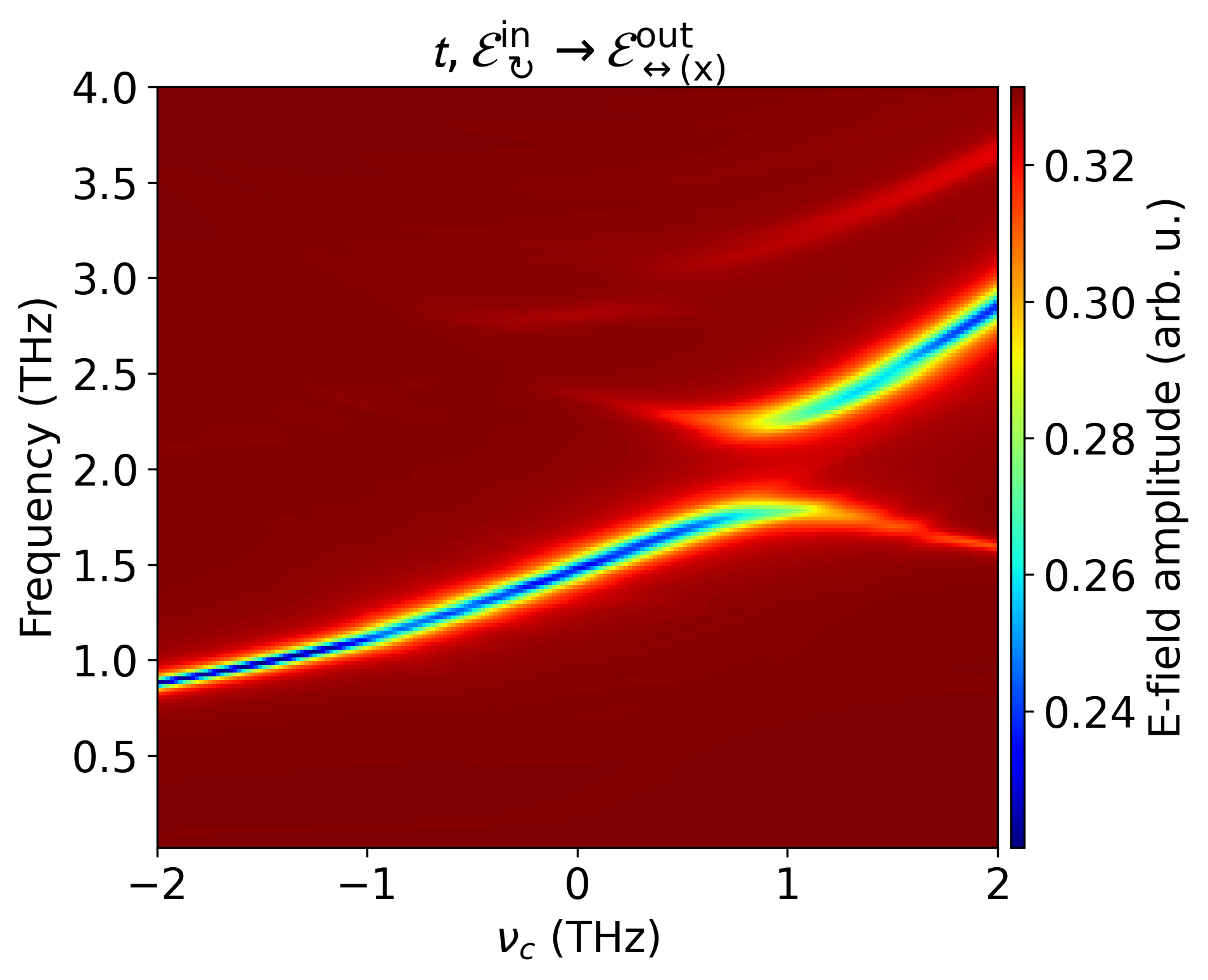}%
    \hspace{-0.01\textwidth}%
    \includegraphics[width=0.33\textwidth]{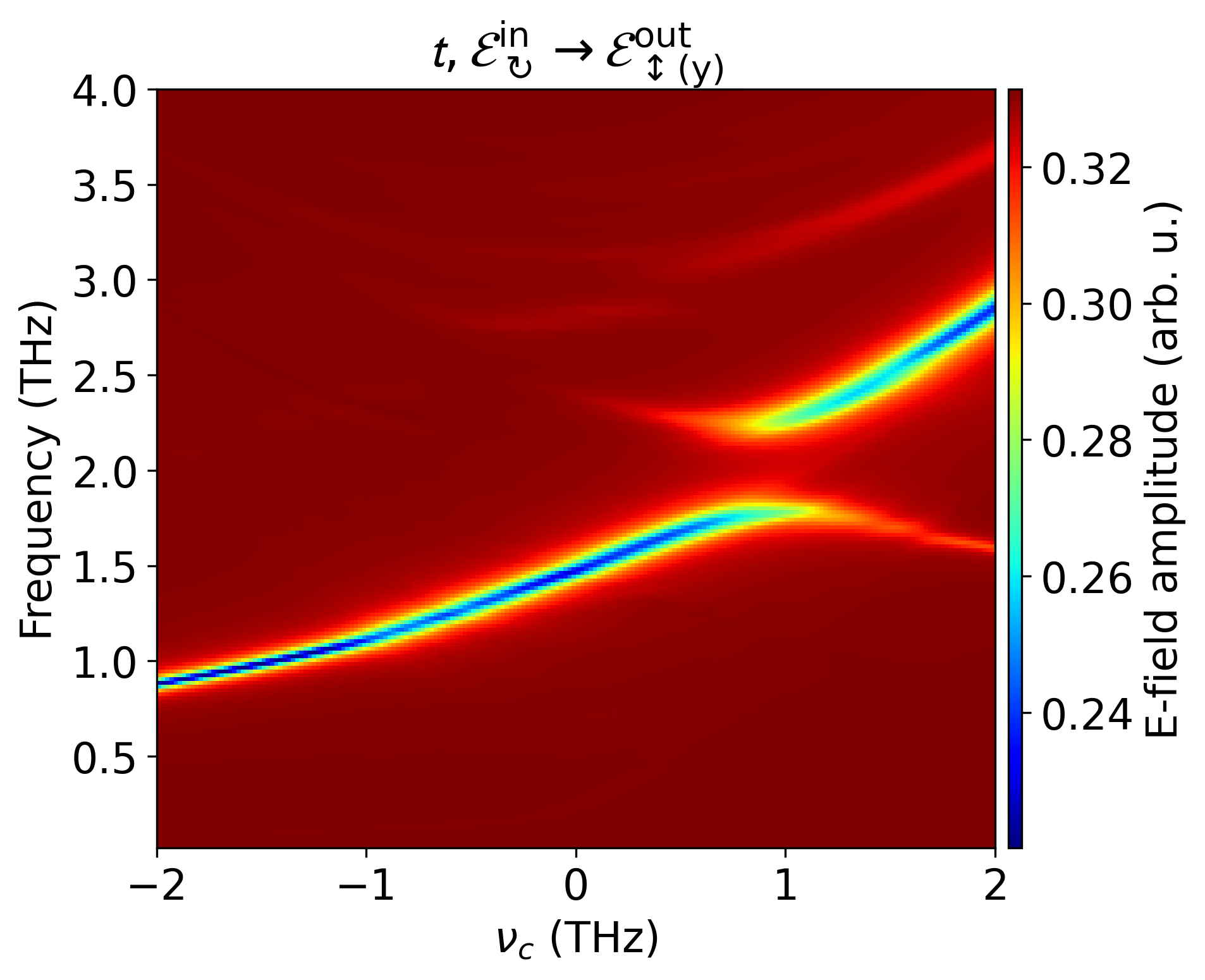}
    \caption{FEM far-field transmission spectra for the square patch with right-circularly polarized incident light. The panels show the transmitted right-circularly polarized (left), linearly $x$-polarized (center) and linearly $y$-polarized field component (right).}
    \label{fig:square}
\end{figure}

\begin{figure}[h!]
    \centering
    \includegraphics[width=0.33\textwidth]{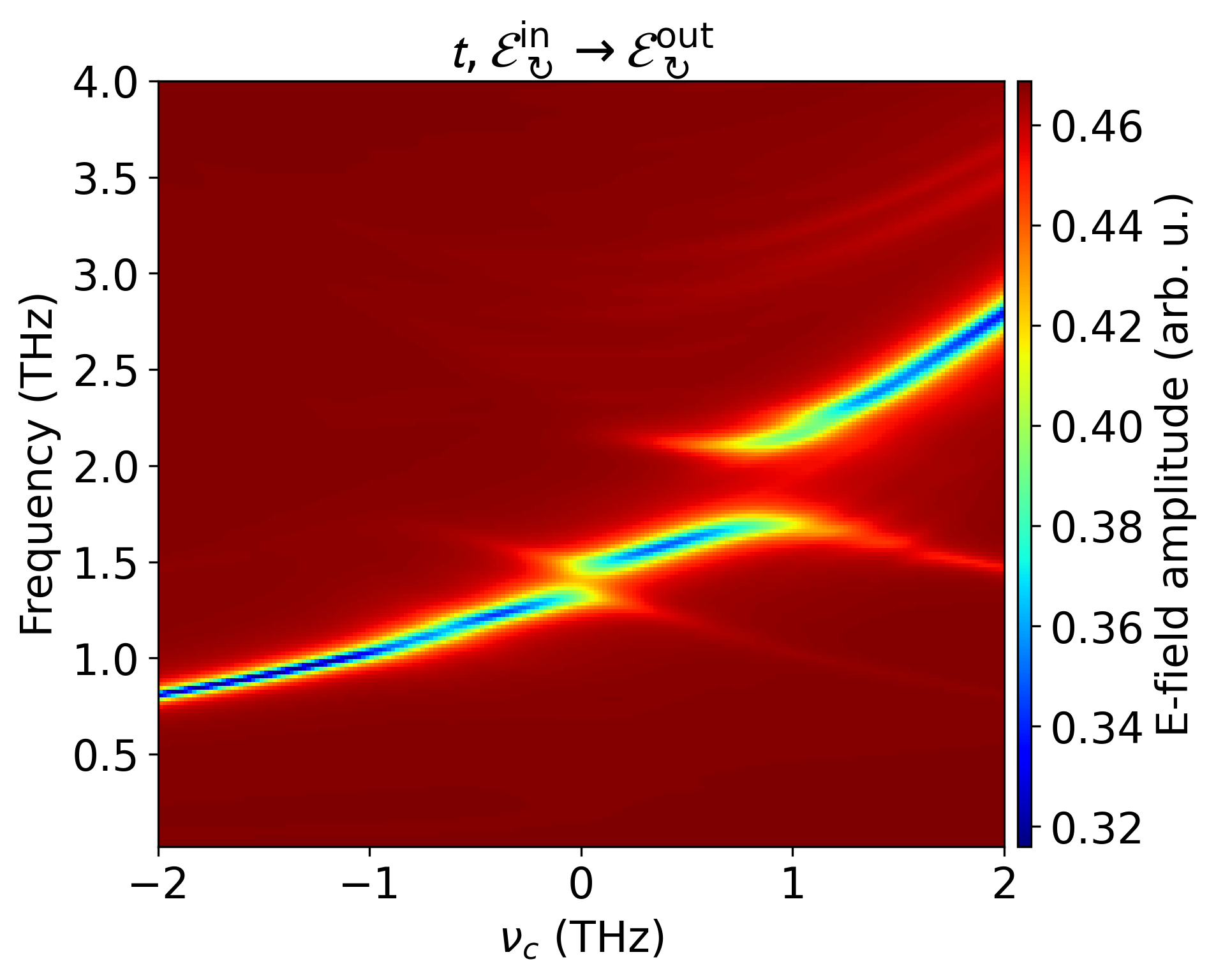}%
    \includegraphics[width=0.33\textwidth]{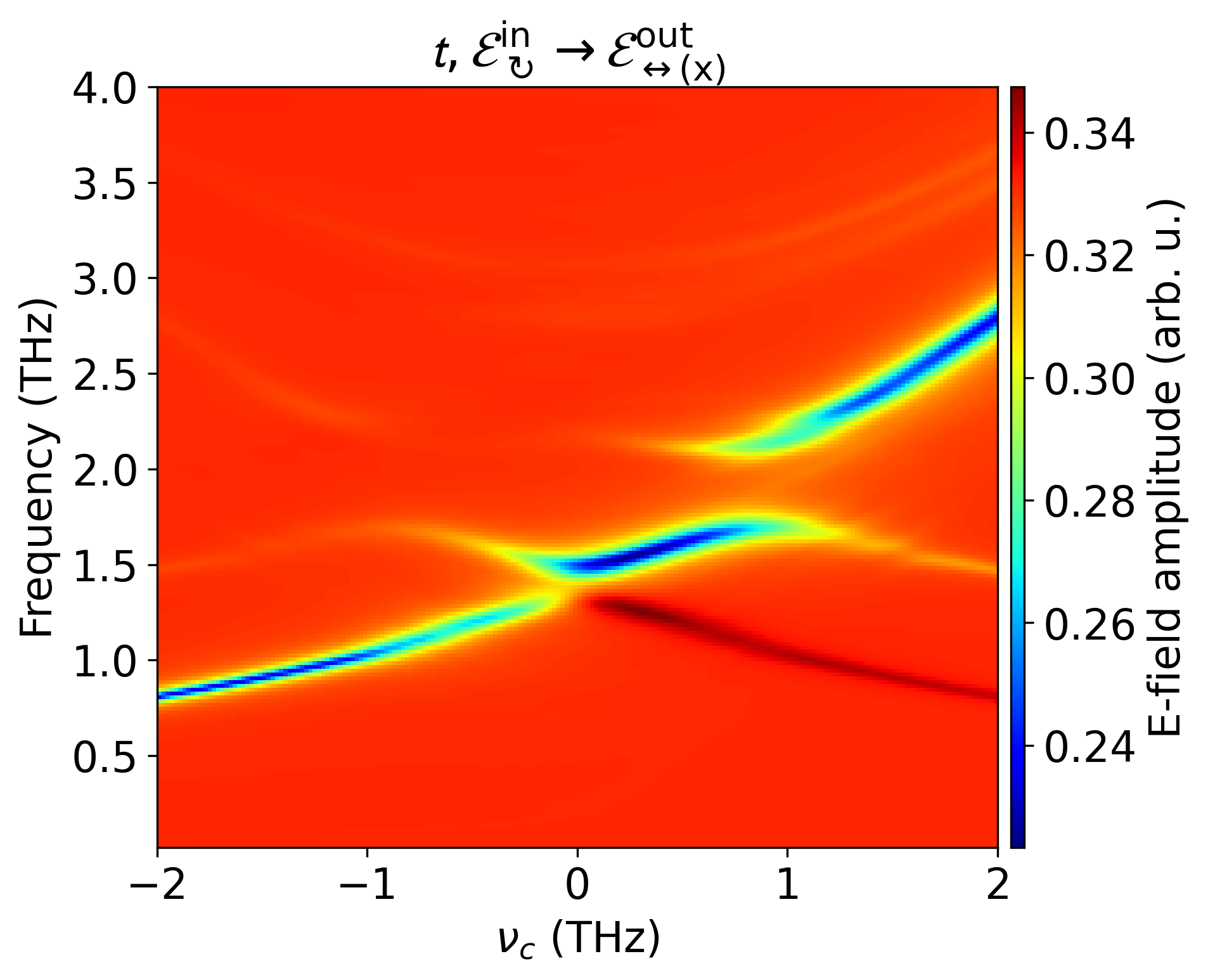}%
    \includegraphics[width=0.33\textwidth]{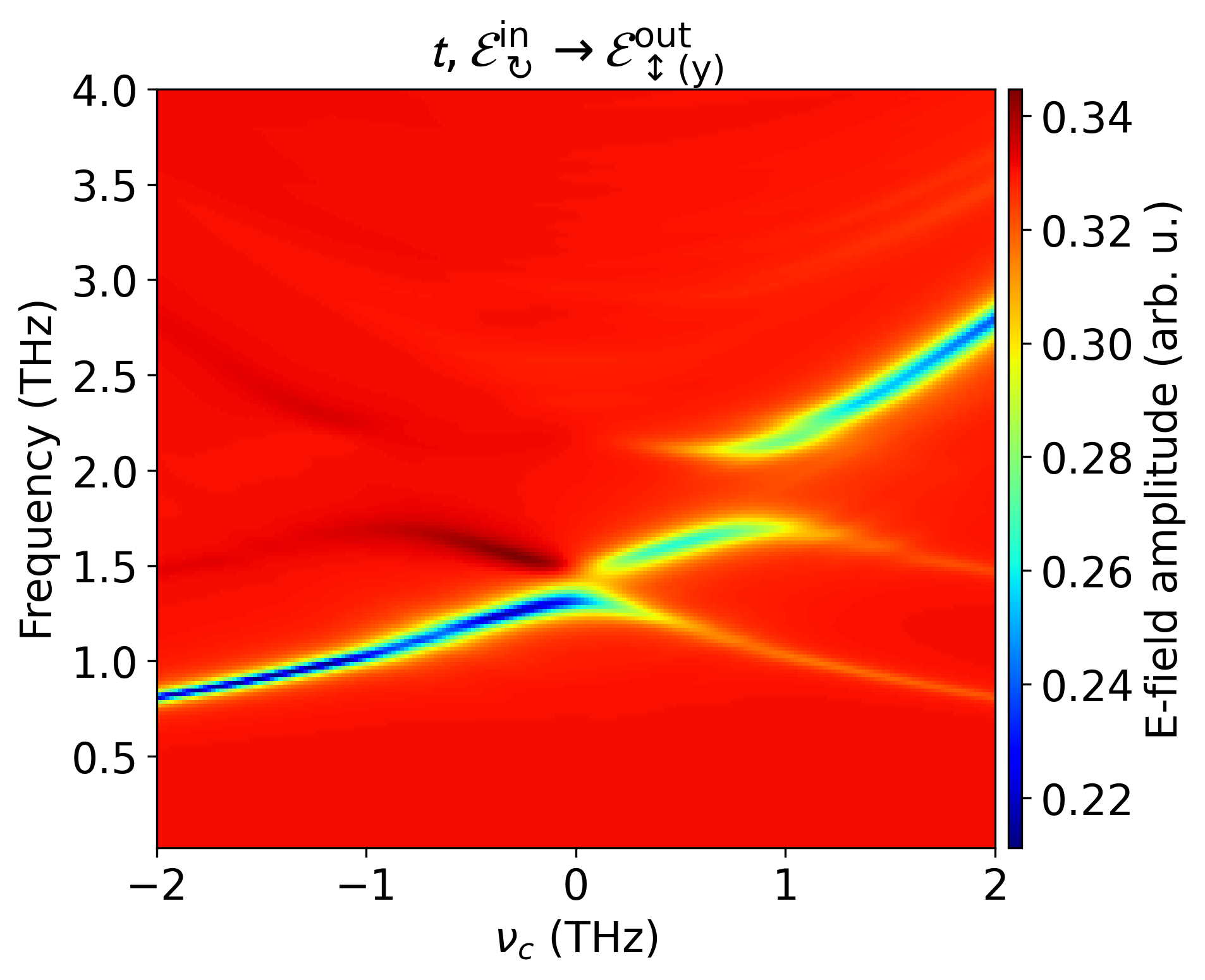}
    \caption{FEM far-field transmission spectra for the rectangular patch with right circularly-polarized incident light. The panels show the transmitted right-circularly polarized (left), linearly $x$-polarized (center) and linearly $y$-polarized field component (right).}
    \label{fig:rect}
\end{figure}

\begin{figure}[h!]
    \centering
    \includegraphics[width=0.33\textwidth]{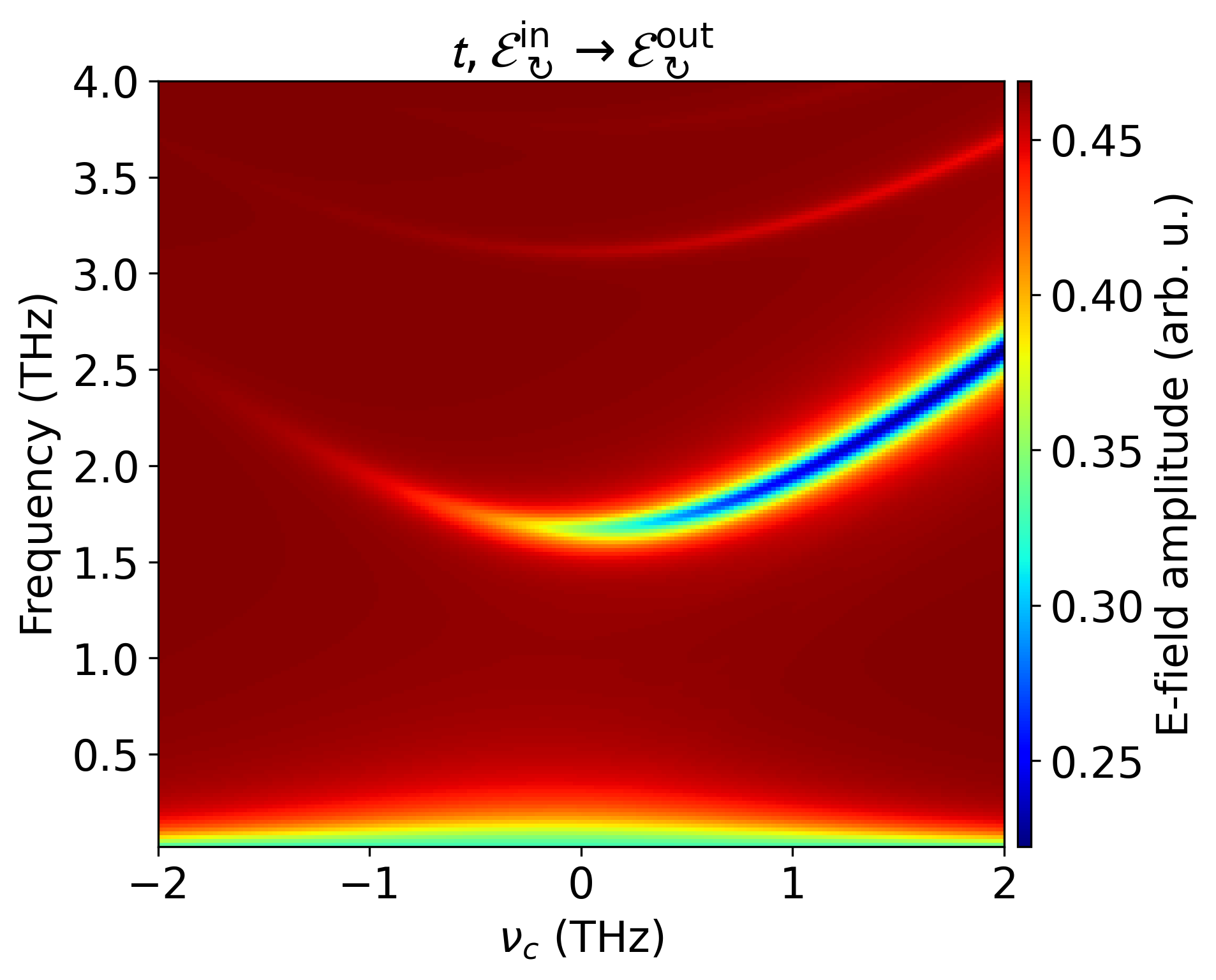}%
    \includegraphics[width=0.33\textwidth]{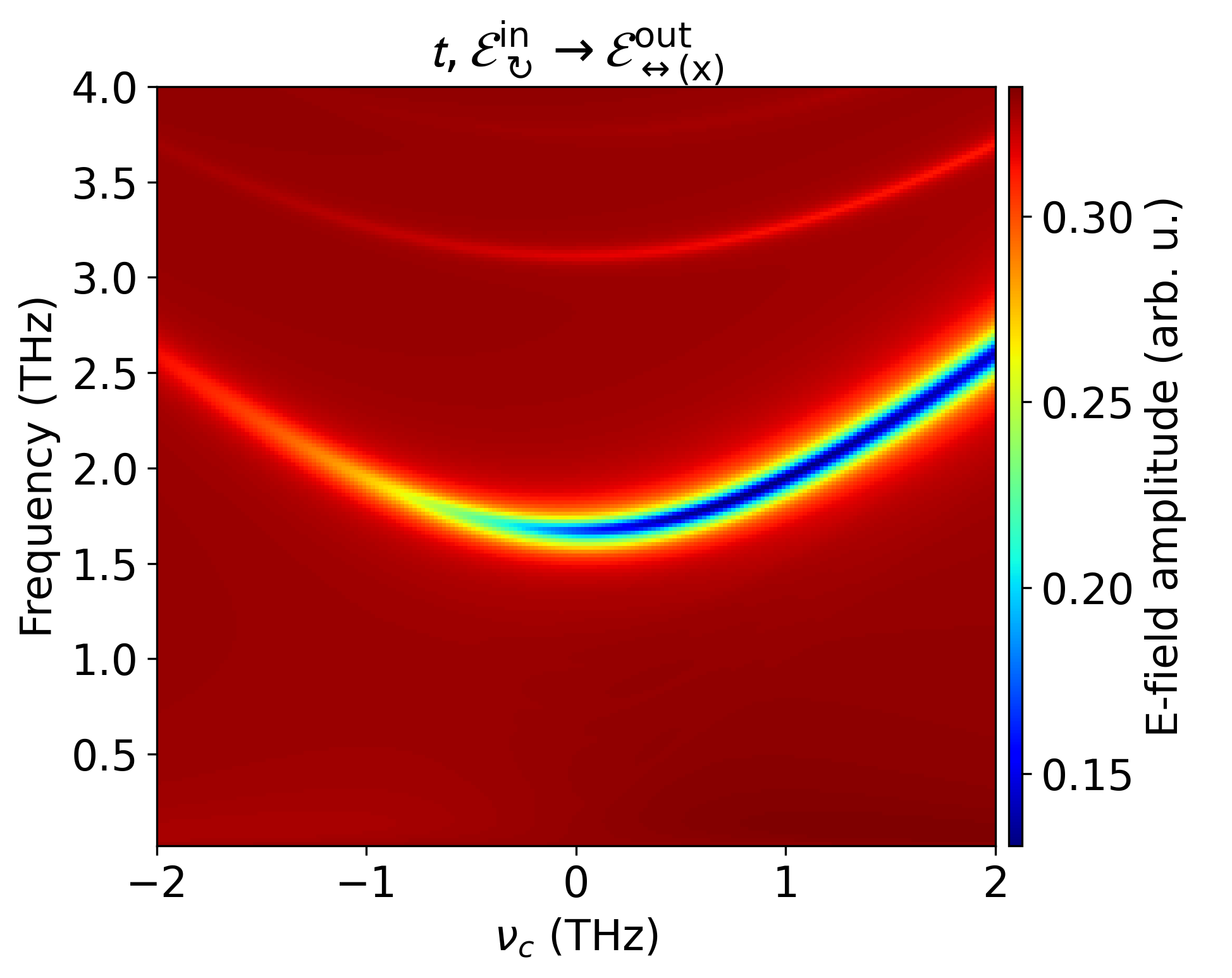}%
    \includegraphics[width=0.33\textwidth]{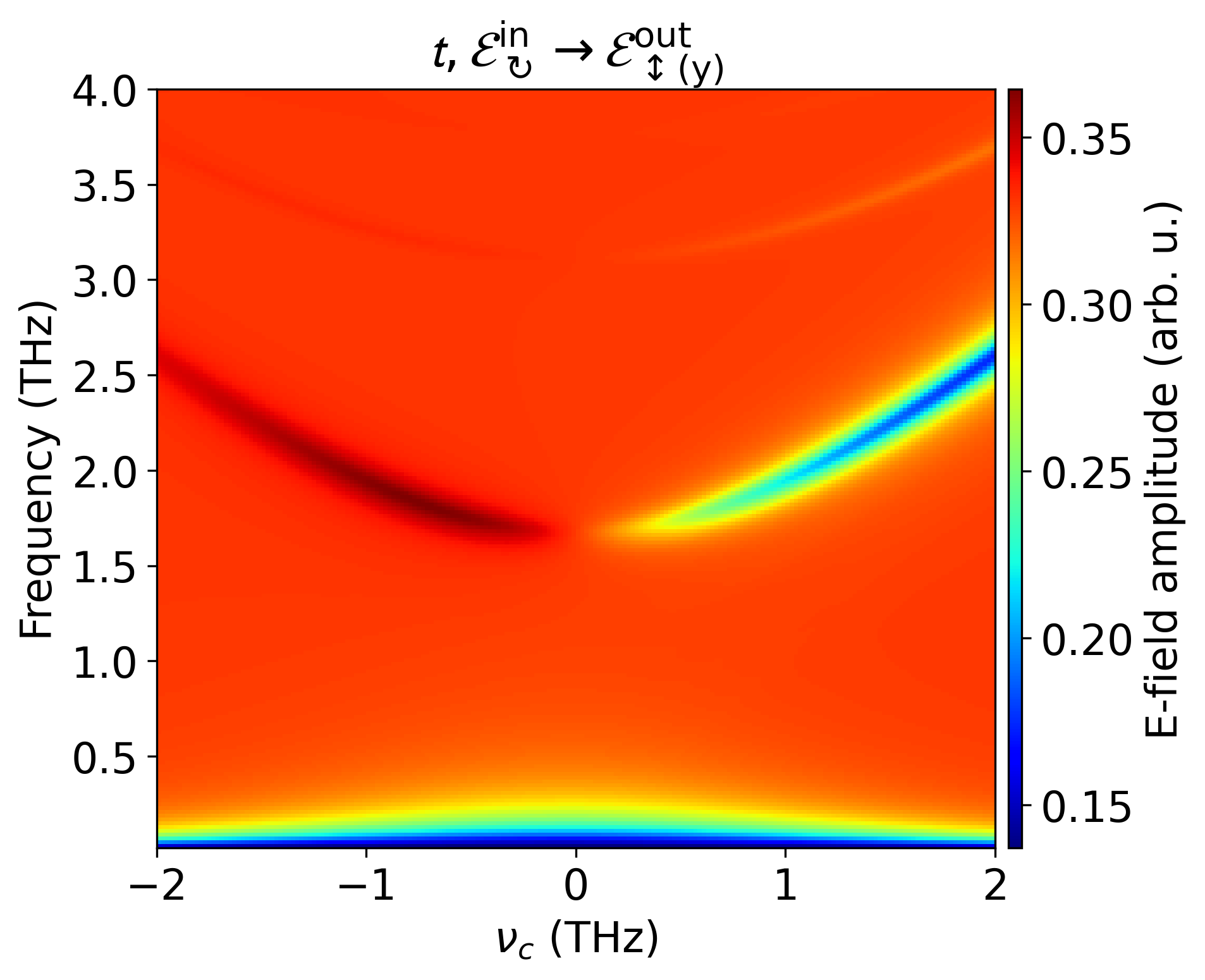}
    \caption{FEM far-field transmission spectra for the stripe with right circularly-polarized incident light. The panels show the transmitted right-circularly polarized (left), linearly $x$-polarized (center) and linearly $y$-polarized field component (right).}
    \label{fig:stripe}
\end{figure}

In support of our theory we have used finite-element frequency domain calculations with a similar approach as the one detailed in \cite{Bayer2017}, where the single quantum well is modeled as an anisotropic effective medium. The simulation volume is set to 40\,$\mu$m in the $z$-direction in order to extract the proper far-field solutions from the bottom plane of volume with scattering boundary conditions. In the the $xy$-plane it is instead set to two times the dimension of the patches. The different polarization combinations used as in the simulations of Fig.\,4 from the main text follow in Fig.~\ref{fig:square} for the square geometry, Fig.~\ref{fig:rect} for the rectangular geometry, and Fig.~\ref{fig:stripe} for the stripe.

\newpage

\begin{figure}[h!]
    \centering
    \includegraphics{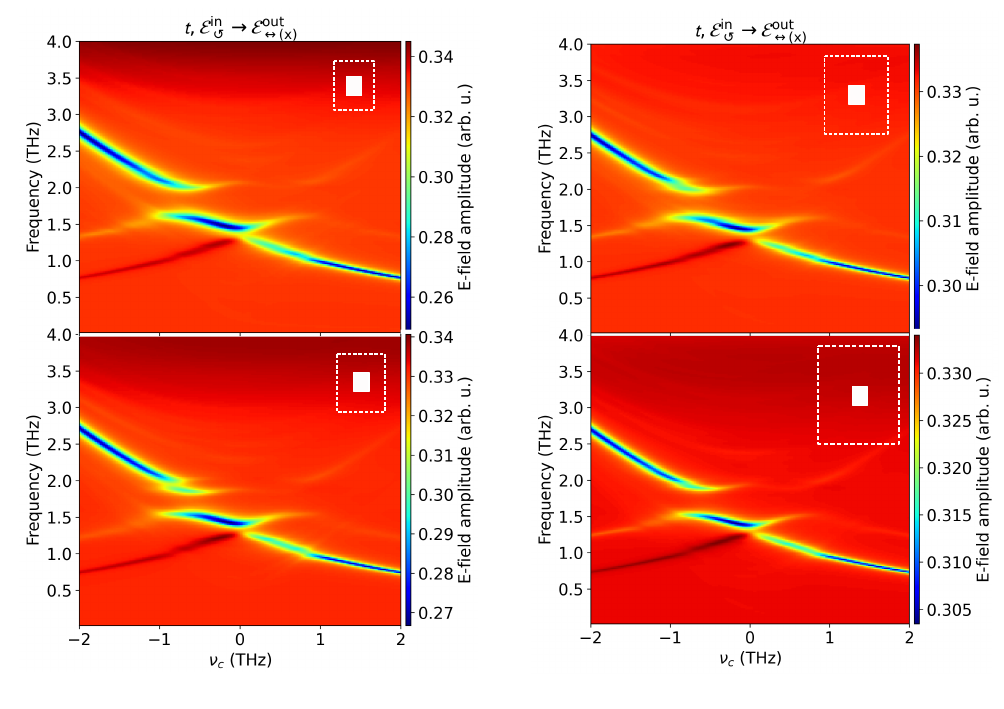}
    \caption{FEM far-field transmission spectra for the rectangular patch with left-circularly polarized incident light for various dimensions of the unit cell. The rectangle patch dimension is $2.8\mu\mathrm{m}\times3.4\,\mu\mathrm{m}$, while the unit cell is, from top left to bottom right, $7\mu\mathrm{m}\times8.5\,\mu\mathrm{m}$, $8.4\,\mu\mathrm{m}\times10.2\,\mu\mathrm{m}$, $11.2\,\mu\mathrm{m}\times13.6\,\mu\mathrm{m}$, and $14\,\mu\mathrm{m}\times17\,\mu\mathrm{m}$, respectively.}
    \label{fig_UC}
\end{figure}

\begin{figure}[h!]
    \centering
    \includegraphics{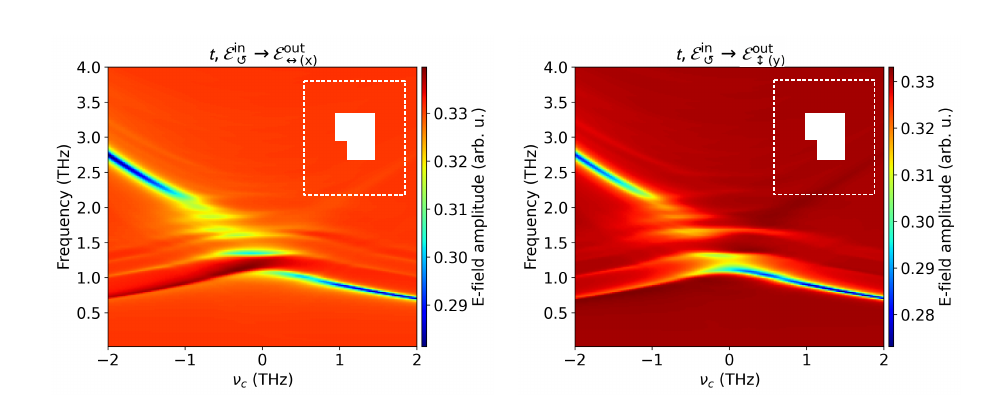}
    \caption{FEM far-field transmission spectra for an L-shaped patch with left-circularly polarized incident light. The geometrical parameters are $3\,\mu\mathrm{m}\times3.5\,\mu \mathrm{m}$ outer dimensions, with a rectangular piece (bottom left) of $1\,\mu\mathrm{m}\times 1.5\mu\mathrm{m}$ removed.}
    \label{fig_Lshape}
\end{figure}

\indent We have also performed a sweep of the unit cell [Fig.\,\ref{fig_UC}] for an individual patch (rectangle, $2.8\,\mu\mathrm{m}\times3.4\,\mu\mathrm{m}$) to prove that the influence of the array periodicity on the coupling is negligible compared to the symmetry class of the geometry. In Fig. \ref{fig_Lshape} we provide a perspective on inherently chiral geometries (in this case an L-shape), which open up additional coupling pathways. The data show an entire fan of coupled MPP modes stemming from hybridization of multiple magnetoplasmon modes in this significantly more complex confinement setting as compared to the basic geometric shapes.

\newpage

\end{document}